\documentclass[preprint,showpacs,preprintnumbers,amsmath,amssymb,prb,aps]{revtex4-1}
\usepackage{epsfig}
\usepackage{graphicx}
\usepackage{dcolumn}
\usepackage{bm}
\usepackage{slashbox}
\usepackage{textcomp}

\begin{document}

\title{Density matrix approach to the orbital ordering in the spinel vanadates: A case study}
\author{Sohan Lal} 
\altaffiliation{Electronic mail:goluthakur2007@gmail.com}
\author{Sudhir K. Pandey}
\affiliation{School of Engineering, Indian Institute of Technology Mandi, Kamand 175005, Himachal Pradesh, India}

\date{\today}

\begin{abstract}

    In this work we apply the density matrices approach to orbital ordering (OO) in order to study the OO of the spinel vanadates AV$_{2}$O$_{4}$ (A $\equiv$ Zn, Cd and Mg), which is normally believed to be responsible for the structural transition from cubic to tetragonal phase observed in these compounds. The density matrices of vanadium atoms are obtained by using {\it state-of-the-art} full-potential linearized augmented plane wave method based GGA+U calculations. In the absence of spin-orbit coupling, the present study shows the existence of anti-ferro OO in the global (local octahedral) coordinate system where $d_{xz}$ and $d_{yz}$ ($d_{xz}$+$d_{yz}$ and $d_{xz}$-$d_{yz}$) orbitals are mainly occupied at the neighboring V sites for all the compounds.
            
\end{abstract}


\maketitle

\section{Introduction} 
  
    The orbital degrees of freedom play important role in the condensed matter physics to understand the various phenomena observed in the strongly correlated systems from last few decades\cite{Khomskii,Imada,Tokura,Dagotto,Hotta}. Transition metals having partially filled degenerate ${d}$ orbitals, where the certain kind of orbital being occupied at specific site is expected to be directed by the another kind of orbital being occupied at another site. Hence such type of preferred  occupation of orbitals in a system can lead to the various kind of orbital ordering (OO). The OO is strongly linked with the spin ordering (SO) in many compounds. In order to understand the various SO observed in La$_{1-x}$Ca$_x$MnO$_3$ compounds Goodenough has proposed different kinds of OO\cite{Goodenough}. The famous Goodenough-Kanamori-Anderson rules provide the scheme which help in predicting the possible OO for a given SO observed in a compound\cite{Khomskii2010}. In transition metal oxides Coulomb repulsion {\it U} are found to play an important role in driving the OO. However it is still unclear whether {\it U} only enhances the effect of lattice distortions or really drives the OO in the transition metal oxides via super-exchange mechanism\cite{Pavarini,Koch}. OO is also responsible for the structural transition as proposed in the spinel vanadates\cite{Huang,Radaelli}. According to Tchernyshov the structural transition arises in the spinels from cubic to tetragonal due to the interplay of the Jahn-Teller effect and relativistic spin-orbit interaction\cite{Tchernyshyov}. Khomskii {\it et al.} suggested a model that predicts the structural ground state due to the anti-Jahn Teller effect caused by the broadening of the $yz$, $xz$ bands and leads to the orbitally driven Peierls state\cite{Mizokawa}. However Tsunetsugu {\it et al.} showed that the Coulomb and exchange interaction between the magnetic ions and the coupling to the Jahn-Teller lattice distortion are responsible for the structural transition in the spinels\cite{Motome}. OO also affects the transport behavior of a compound because of the localization of the electrons in the orbitals. The occupancy of orbitals and their orientation in the space leads the anisotropic behavior in the system\cite{Kugel,Kugel1975,Khomskii1982}.
   
  Before 1990's {\it ab initio } density functional theory was not a popular approach for calculating the OO in a strongly correlated system. For most of the materials the OO were calculated by using the model calculations\cite{Kugel1973}. However with the implementation of LDA+{\it U} scheme in the {\it ab initio } density functional theory based codes the study of OO in many materials became popular. In this approach the density matrices (DM) of the correlated orbitals are the natural output of the LDA+{\it U} calculations. These DM are used for knowing the orbital occupancies of orbitals. Most of the codes calculates the DM in the spherical harmonic representation. Normally the OO is shown in cubic harmonic representation. Moreover, knowing the OO in cubic harmonics looking at the DM in the spherical harmonics is not a straightforward job. This becomes more complex if the transition metal is not in the regular octahedral symmetry. 

   In this work we use the DM of vanadium atoms to study the OO pattern of the spinel vanadates, which has been a contentious issue for a long time\cite{Tsunetsugu,Lee,Tchernyshyov,Matteo,Mizokawa,Maitra,Pandey2011,Pandey2012,Wheeler}. Spinel vanadates with general formula AV$_{2}$O$_{4}$ (A $\equiv$ Zn, Cd and Mg) possessing cubic structure at room temperature and forming pyrochlore lattice at the V sites, which is responsible for the geometrical frustration\cite{Maitra,Lee,Onoda,Reehuis,Wheeler,Mamiya,Nishiguchi,Radaelli,Tsunetsugu,Tchernyshyov,Matteo}. At low temperature these compounds show structural transition from cubic to tetragonal\cite{Onoda,Reehuis,Wheeler,Mamiya,Nishiguchi,Radaelli}. The driving force for the structural transition is proposed to be the OO\cite{Roger}. Different studies show different OO in these compounds. For example, based on model calculations, Motome {\it et al.} have predicted the anti-ferro OO, where $d_{xz}$ and $d_{yz}$ orbitals are mainly occupied alternatively along c-axis together with the occupancy of $d_{xy}$ orbital at every V sites\cite{Tsunetsugu}. This OO is inconsistent with the spatial symmetry {\it I}4$_{1}$/{\it amd} of tetragonal phase as it breaks the mirror reflections in the planes (110) and (1$\bar{1}$0) and diamond glides {\it d} in the planes (100) and (010)\cite{Reehuis,Nishiguchi}. However, in the presence of strong relativistic spin-orbit coupling (SOC) and inclusion of Jahn-Teller effect as a perturbation the model calculations of Tchernychyov have predicted a complex uniform ferro-orbital ordering of type $d_{xz}$$\pm$i$d_{yz}$ at each V sites\cite{Tchernyshyov}. This OO is consistent with the spatial symmetry {\it I}4$_{1}$/{\it amd} as it preserves both mirror and glide symmetries. Matteo {\it et al.} proposed a theoretical model in which the spin-orbital super-exchange interaction and SOC have been treated on equal footing. They have shown the existence of complex orbital ordered ground state similar to that of Tchernychyov for intermediate spin-orbit coupling strength with the symmetry {\it I}4$_{1}$/{\it amd}\cite{Matteo}. Khomskii {\it et al.} proposed a mechanism considering the superstructure formed in spinel containing transition metal ions. They have shown that close to the itinerant state the orbital-Peierls ordering of type $d_{yz}$-$d_{yz}$-$d_{xz}$-$d_{xz}$ along the tetragonally compressed c-axis have been observed\cite{Mizokawa}. Based on 3-dimensional electron spin-density plots obtained from LSDA+{\it U} calculations on ZnV$_2$O$_4$ compound, Maitra {\it et al.} have shown the existence of $d_{xz}$+$d_{yz}$ and $d_{xz}$-$d_{yz}$ OO in the absence of SOC. In the presence of SOC they found the mixture of $d_{xz}$+i$d_{yz}$ and $d_{xz}$-i$d_{yz}$ orbitals\cite{Maitra}. 
    
  In order to study the OO of AV$_{2}$O$_{4}$ (A $\equiv$ Zn, Cd and Mg) compounds, we use the DM (in the spherical harmonics representation) of vanadium atoms obtained from GGA+{\it U} calculations. Our studies in the absence of SOC suggest the anti-ferro OO in the global (local octahedral) coordinate system for all the compounds where $d_{xz}$ and $d_{yz}$ ($d_{xz}$+$d_{yz}$ and $d_{xz}$-$d_{yz}$) orbitals are mainly occupied at the neighboring V sites.   
      
\section{Computational Detail}       
    
    The DM of vanadium atoms in ZnV$_{2}$O$_{4}$, MgV$_{2}$O$_{4}$ and CdV$_{2}$O$_{4}$ compounds are calculated by using the {\it state-of-the-art} full-potential linearized augmented plane wave (FP-
LAPW) method \cite{elk}. The lattice parameters and atomic positions used here for every compounds are taken from the literature\cite{Onoda,Reehuis,Wheeler}. PBEsol exchange correlation functional is used in the calculation \cite{Perdew}.  The effect of on-site Coulomb interaction among V 3d electrons is considered within GGA+{\it U} formulation of the density functional theory \cite{Bultmark}. Fully localized double counting scheme has been considered in the calculations\cite{Liechtenstein}. In this method {\it U} and {\it J} are used as parameters. We have varied {\it U} from 3 eV to 5 eV at fixed {\it J} = 0.5 eV and obtained the DM of V atoms in these compounds which are shown in the Tables 1\textendash9.
One can see a slight change in the values of the DM when {\it U} increases from 3 to 4 eV. It remains almost the same for {\it U} = 4-5 eV. In rest of the paper we present the results corresponds to {\it U} = 4 eV.  These value of {\it U} and {\it J} are also in consistent with our earlier work\cite{Pandey2011,Pandey2012}. DM for these values of {\it U} and {\it J} are calculated within the muffin-tin sphere of V atoms. The  muffin-tin sphere radii used in the calculations are 2.0, 2.0, 1.5, 1.8 and 1.5 Bohr for Zn, Cd, Mg, V and O, respectively. (6,6,6) k-point mesh size are used. Convergence target of total energy was achieved below ~10$^{-4}$ Hartrees/cell. It is important to note that the values of DM show small dependence on the size of the muffin-tin radii used in the calculation. However the small variation in the DM due to change in the values of $U$ and muffin-tin radii is not expected to change in the OO pattern obtained in the present work.

\section{Result and Discussion} 

   In the context of geometrical frustration the basic structural unit of the spinel vanadates is the tetrahedron shown in Fig. 1, where four V atoms represented by V1, V2, V3 and V4 are sitting at the four corners of the tetrahedron. In order to know the orbital occupancies of these V atoms we have calculated the DM of these atoms given in spherical harmonic  representation. The DM of V1 and V3 (V2 and V4) are found to be the same suggesting the same orbital occupancies for V1 and V3 (V2 and V4) atoms. If one compares the DM of V1 and V2 atoms one can see the obvious differences in the up-spin channel as given in Tables 1\textendash3 for ZnV$_2$O$_4$. The DM for the combination of $m1$, $m2$ element having values ($\pm$2,0),  (0,$\pm$2), (-1,1), (1,-1) of V$_{1}$ and V$_{2}$ (in brackets) atoms are different. For example (-1,1) element of DM for V1 and V2 atoms are 0.32 and -0.32, respectively. This suggests the different orbital occupancies for V1 and V2 atoms. This is also valid for MgV$_2$O$_4$ and CdV$_2$O$_4$ compounds as evident from Tables 4\textendash9. The DM for the combination of $m1$, $m2$ element having values (-1,1), (1,-1) of V$_{1}$ and V$_{2}$ (in brackets) atoms for MgV$_2$O$_4$ and ($\pm$2,0), (0,$\pm$2), (-1,1), (1,-1) of V$_{1}$ and V$_{2}$ (in brackets) atoms for CdV$_2$O$_4$ compounds are different. This also suggests the different orbital occupancies for V1 and V2 atoms. From the above observations one can conclude the different orbital occupancies for V1 and V2 atoms for all the three compounds. Normally, the OO is represented in the cubic harmonics. In order to know the orbital occupancies of V atoms in these compounds, we have considered the only those elements of the DM whose magnitude is $\geq$ 0.1. From the given DM in Table 1 it is not obvious to judge which orbitals are occupied in the cubic harmonic representation. In order to get the exact orbital occupancies of the strongly correlated electron system in the cubic harmonics for the V1 and V2 atoms we have used the following method:

     The electronic charge distribution of the strongly correlated electron systems in the spherical polar coordinate system can be expressed by using formula, 

\begin{eqnarray}
n(\theta,\phi)\approx \Sigma_{\sigma}\Sigma_{\ell}\Sigma_{m,m'}n^{\sigma}_{\ell,m,m'}Y^{*}_{\ell,m}(r)Y_{\ell,m'}(r)
\label{eq.1}
\end{eqnarray}

where $Y_{\ell,m}$ and n$^{\sigma}_{\ell,m,m'}$ is a spherical harmonic function and $m$, $m'$ element of the DM, respectively\cite{Kozhevnikov,Shick}. Normally strongly correlated orbitals like 3d and 4f are responsible for the OO in the strongly correlated compounds. For 3$d$ electrons system like spinel vanadates where only up spin electrons are responsible for the OO one can write Eqn. 1 as  
 
\begin{eqnarray}
n(\theta,\phi)=\Sigma_{m,m'}n_{m,m'}Y^{*}_m(r)Y_{m'}(r)
\label{eq.2}
\end{eqnarray}

For {\it d} orbitals there will be five different values for the $Y_m$ corresponding to each values of m\cite{Demtroder}. These are $Y_{\pm2}$ = $\frac{1}{4}$$\sqrt{\frac{15}{2\pi }}$$\sin^{2}\theta e^{\pm2i\phi }$, $Y_{\pm1}$=$\frac{1}{2}$$\sqrt{\frac{15}{2\pi}}$$\sin\theta\cos\theta e^{\pm i\phi}$  and $Y_0$=$\frac{1}{4}$$\sqrt{\frac{5}{\pi}}(3\cos^{2}\theta-1)$. Putting these values in Eqn. 2 and using relations:
$x$ = $r$$\sin\theta$$\cos\phi$, $y$ = $r$$\sin\theta$$\sin\phi$ and $z$ = $r$$\cos\theta$, one can find the electron densities of vanadium atoms in the Cartesian coordinate.\\ 

\underline{\textbf{ZnV$_{2}$O$_{4}$ compound:}}\\

By using the above given procedure we can calculate the electron densities n$_{1}$ and n$_{2}$ of V1 and V2 atoms, respectively in Cartesian coordinate which are 
\begin{align*}
n_{1}&= \frac{1}{(x^{2}+y^{2}+z^{2})^{2}}[0.43{\frac{15}{32\pi}(x^{2}+y^{2})^{2}}+0.28\frac{15}{32\pi}\{(x^{2}-y^{2})^{2}-4x^{2}y^{2}+4ixy(x^{2}-y^{2})\}\\
&+0.41{\frac{15}{8\pi}}(x^{2}+y^{2})z^{2}+0.32\frac{15}{8\pi}\{(x^{2}-y^{2})z^{2}+2ixyz^{2}\}+0.17\frac{5}{16\pi}\{2z^{2}-(x^{2}+y^{2})\}^{2}\\
&+0.32\frac{15}{8\pi}\{(x^{2}-y^{2})z^{2}-2ixyz^{2}\}+0.41\frac{15}{8\pi}(x^{2}+y^{2})z^{2} \\
&+0.28\frac{15}{32\pi}\{(x^{2}-y^{2})^{2}-4x^{2}y^{2}-4ixy(x^{2}-y^{2})\}+0.43\frac{15}{32\pi}(x^{2}+y^{2})^{2}]\\
&= \frac{15}{32\pi(x^{2}+y^{2}+z^{2})^{2}}[0.86(x^{4}+y^{4}+2x^{2}y^{2})+0.56(x^{4}+y^{4}-6x^{2}y^{2})+3.28(x^{2}+y^{2})z^{2}\\
&-2.56(y^{2}-x^{2})z^{2}+0.44z^{4}+0.11(x^{4}+y^{4})+0.22x^{2}y^{2}-0.44(x^{2}+y^{2})z^{2}]\\
&= \frac{0.15}{(x^{2}+y^{2}+z^{2})^{2}}[1.53(x^{4}+y^{4})-1.42x^{2}y^{2}+0.28(x^{2}+y^{2})z^{2}+0.44z^{4}+5.12x^{2}z^{2}]\\
\end{align*}

Thus

\begin{eqnarray}
n_{1}=\frac{1}{(x^{2}+y^{2}+z^{2})^{2}}[0.23(x^{4}+y^{4})-0.21(xy)^{2}+0.04(x^{2}+y^{2})z^{2}+0.07z^{4}
+0.77(xz)^{2}]
\end{eqnarray}

and

\begin{align*}
n_{2}&= \frac{1}{(x^{2}+y^{2}+z^{2})^{2}}[0.43{\frac{15}{32\pi}(x^{2}+y^{2})^{2}}+0.28\frac{15}{32\pi}\{(x^{2}-y^{2})^{2}-4(xy)^{2}+4ixy(x^{2}-y^{2})\}\\
&+0.41{\frac{15}{8\pi}}(x^{2}+y^{2})z^{2}-0.32\frac{15}{8\pi}\{(x^{2}-y^{2})z^{2}+2ixyz^{2}\}+0.17\frac{5}{16\pi}\{2z^{2}-(x^{2}+y^{2})\}^{2}\\
&-0.32\frac{15}{8\pi}\{(x^{2}-y^{2})z^{2}-2ixyz^{2}\}+0.41\frac{15}{8\pi}(x^{2}+y^{2})z^{2}\\
&+0.28\frac{15}{32\pi}\{(x^{2}-y^{2})^{2}-4(xy)^{2}-4ixy(x^{2}-y^{2})\}+0.43\frac{15}{32\pi}(x^{2}+y^{2})^{2}]\\
&= \frac{15}{32\pi(x^{2}+y^{2}+z^{2})^{2}}[0.86(x^{4}+y^{4}+2x^{2}y^{2})+0.56(x^{4}+y^{4}-6x^{2}y^{2})+3.28(x^{2}+y^{2})z^{2}\\
&-2.56(x^{2}-y^{2})z^{2}+0.44z^{4}+0.11(x^{4}+y^{4})+0.22x^{2}y^{2}-0.44(x^{2}+y^{2})z^{2}]\\
&= \frac{0.15}{(x^{2}+y^{2}+z^{2})^{2}}[1.53(x^{4}+y^{4})-1.42x^{2}y^{2}+0.28(x^{2}+y^{2})z^{2}+0.44z^{4}+5.12y^{2}z^{2}]\\
\end{align*}

Thus 

\begin{eqnarray}
n_{2}=\frac{1}{(x^{2}+y^{2}+z^{2})^{2}}[0.23(x^{4}+y^{4})-0.21(xy)^{2}+0.04(x^{2}+y^{2})z^{2}+0.07z^{4}
+0.77(yz)^{2}]
\end{eqnarray}

Eqns. 3 and 4 provides the spatial distribution of the electron density in the Cartesian coordinate which gives the information about the OO in ZnV$_2$O$_4$ compound. The first four terms of Eqns. (3) and (4) are same and hence contribute equal electron density in the space for both the atoms. The last term is different and can be considered as a signature of OO as for V1 atom electrons are mainly occupied in the ${xz}$ plane whereas for the V2 atom electrons are mainly occupied in ${yz}$ plane. This suggests that the $d_{xz}$ and $d_{yz}$ orbitals are mainly occupied by V1 and V2 atoms, respectively which corresponds to the anti-ferro OO in ZnV$_2$O$_4$. 

The OO of ZnV2O4 is studied by many researchers. It will be interesting to compare our result with the results of other groups. Based on model calculations, Tsunetsugu \textit{et al.} have predicted $d_{xz}$ and $d_{yz}$ OO. However, the \emph{ab inito} electronic structure calculations of Maitra \textit{et al.} have shown the existence of $d_{xz}$+$d_{yz}$ and $d_{xz}$-$d_{yz}$ OO. It is important to note that coordinate systems used by these groups are different from our coordinate system. These groups has set $x$ and $y$ axes along the two basal V-O bonds and $z$ axis along the apical V-O bond of regular VO$_6$ octahedron. In our case the $x$ and $y$ axes are making angles of 47.4 degree and 42.6 degree with two basal V-O bonds and $z$ axis is making an angle of 4.8 degree with the apical V-O bond. This is due to the fact that in tetragonal phase the VO$_6$ octahedron is distorted one. In order to compare our result with the results of Tsunetsugu \textit{et al.} and Maitra \textit{et al.} we have rotated the $x$ and $y$ axes by 42.6 degree and calculated the electron densities for V1 and V2 atoms in new coordinate system. We get
 
\begin{align*}
n_{1}&= \frac{1}{(x^{2}+y^{2}+z^{2})^{2}}[0.43\frac{15}{32\pi}(x^{2}+y^{2})^{2}+0.28\frac{15}{32\pi}\{3.95(xy)^{2}-0.99(y^{2}-x^{2})^{2}-0.64xy(y^{2}-x^{2})\\
&+4i\{0.16(xy)^{2}-0.04(y^{2}-x^{2})^{2}+0.99xy(y^{2}-x^{2})\}\}+0.41\frac{15}{8\pi}(x^{2}+y^{2})z^{2}\\
&+0.32\frac{15}{8\pi}\{1.99xyz^{2}-0.08(y^{2}-x^{2})z^{2}+2i\{0.50(y^{2}-x^{2})z^{2}+0.08xyz^{2}\}\}\\
&+0.17\frac{5}{16\pi}\{2z^{2}-(x^{2}+y^{2})\}^{2}+0.32\frac{15}{8\pi}\{1.99xyz^{2}-0.08(y^{2}-x^{2})z^{2}-2i\{0.50(y^{2}-x^{2})z^{2}\\
&+0.08xyz^{2}\}\}+0.41\frac{15}{8\pi}(x^{2}+y^{2})z^{2}+0.28\frac{15}{32\pi}\{3.95(xy)^{2}-0.99(y^{2}-x^{2})^{2}-0.64xy(y^{2}-x^{2})\\
&-4i\{0.16(xy)^{2}-0.04(y^{2}-x^{2})^{2}+0.99xy(y^{2}-x^{2})\}\}+0.43\frac{15}{32\pi}(x^{2}+y^{2})^{2}]\\
&= \frac{15}{32\pi(x^{2}+y^{2}+z^{2})^{2}}[0.86(x^{4}+y^{4}+2x^{2}y^{2})+2.2x^{2}y^{2}-0.55(x^{4}+y^{4}-2x^{2}y^{2})\\
&-0.35xy(y^{2}-x^{2})+3.28(x^{2}+y^{2})z^{2}+5.09xyz^{2}-0.20(y^{2}-x^{2})z^{2}+0.44z^{4}\\
&+0.11(x^{4}+y^{4})+0.22x^{2}y^{2}-0.44(x^{2}+y^{2})z^{2}]\\
&= \frac{0.15}{(x^{2}+y^{2}+z^{2})^{2}}[0.42(x^{4}+y^{4})+5.24x^{2}y^{2}+0.44z^{4}-0.35xy(y^{2}-x^{2})+0.40x^{2}z^{2}\\
&+2.64(x^{2}+y^{2})z^{2}+5.09xyz^{2}]\\
\end{align*}

Thus

\begin{eqnarray}
n_{1}&=&\frac{1}{(x^{2}+y^{2}+z^{2})^{2}}[0.06(x^{4}+y^{4})+0.79(xy)^{2}+0.07z^{4}-0.05xy(y^{2}-x^{2}) \nonumber \\
&&+0.06(xz)^{2}+0.39(xz+yz)^{2}-0.02xyz^{2}] 
\end{eqnarray}

and 

\begin{align*}
n_{2}&=\frac{1}{(x^{2}+y^{2}+z^{2})^{2}}[0.43\frac{15}{32\pi}(x^{2}+y^{2})^{2}+0.28\frac{15}{32\pi}\{3.95(xy)^{2}-0.99(y^{2}-x^{2})^{2}-0.64xy(y^{2}-x^{2})\\
&+4i\{0.16(xy)^{2}-0.04(y^{2}-x^{2})^{2}+0.99xy(y^{2}-x^{2})\}\}+0.41\frac{15}{8\pi}(x^{2}+y^{2})z^{2}\\
&-0.32\frac{15}{8\pi}\{1.99xyz^{2}-0.08(y^{2}-x^{2})z^{2}+2i\{0.50(y^{2}-x^{2})z^{2}+0.08xyz^{2}\}\}\\
&+0.17\frac{5}{16\pi}\{2z^{2}-(x^{2}+y^{2})\}^{2}-0.32\frac{15}{8\pi}\{1.99xyz^{2}-0.08(y^{2}-x^{2})z^{2}-2i\{0.50(y^{2}-x^{2})z^{2}\\
&+0.08xyz^{2}\}\}+0.41\frac{15}{8\pi}(x^{2}+y^{2})z^{2}+0.28\frac{15}{32\pi}\{3.95(xy)^{2}-0.99(y^{2}-x^{2})^{2}-0.64xy(y^{2}-x^{2})\\
&-4i\{0.16(xy)^{2}-0.04(y^{2}-x^{2})^{2}+0.99xy(y^{2}-x^{2})\}\}+0.43\frac{15}{32\pi}(x^{2}+y^{2})^{2}]\\
&= \frac{15}{32\pi(x^{2}+y^{2}+z^{2})^{2}}[0.86(x^{4}+y^{4}+2x^{2}y^{2})+2.2x^{2}y^{2}-0.55(x^{4}+y^{4}-2x^{2}y^{2})\\
&-0.35xy(y^{2}-x^{2})+3.28(x^{2}+y^{2})z^{2}-5.09xyz^{2}-0.20(x^{2}-y^{2})z^{2}+0.44z^{4}\\
&+0.11(x^{4}+y^{4})+0.22x^{2}y^{2}-0.44(x^{2}+y^{2})z^{2}]\\
&= \frac{0.15}{(x^{2}+y^{2}+z^{2})^{2}}[0.42(x^{4}+y^{4})+5.24x^{2}y^{2}+0.44z^{4}-0.35xy(y^{2}-x^{2})+0.40y^{2}z^{2}\\
&+2.64(x^{2}+y^{2})z^{2}-5.09xyz^{2}]\\
\end{align*}

Thus

\begin{eqnarray}
n_{2}&=&\frac{1}{(x^{2}+y^{2}+z^{2})^{2}}[0.06(x^{4}+y^{4})+0.79(xy)^{2}+0.07z^{4}-0.05xy(y^{2}-x^{2}) \nonumber \\
&&+0.06(yz)^{2}+0.39(xz-yz)^{2}+0.02xyz^{2}] 
\end{eqnarray}

The first four terms of Eqns. (5) and (6) are same and suggesting the same electron density in the space for both the atoms. In order to find the OO pattern we have to look at the terms which are different in the above equations. On dropping the first four terms from Eqns. (5) and (6) we get

\begin{eqnarray}
n_{1}=\frac{1}{(x^{2}+y^{2}+z^{2})^{2}}[0.39(xz+yz)^{2}+0.06(xz)^{2}-0.02xyz^{2}]
\end{eqnarray}
and
\begin{eqnarray}
n_{2}=\frac{1}{(x^{2}+y^{2}+z^{2})^{2}}[0.39(xz-yz)^{2}+0.06(yz)^{2}+0.02xyz^{2}]   
\end{eqnarray}          
                                   
On neglecting the last two terms which are very small in comparison to first term, Eqns. (7) and (8) clearly indicates that the electron densities of both V1 and V2 atoms are mainly in $xz$+$yz$ and $xz$-$yz$ planes, respectively. Hence $d_{xz}$+$d_{yz}$ and $d_{xz}$-$d_{yz}$ orbitals are mainly occupied at the neighboring sites. This type of orbital occupancies corresponds the anti-ferro OO in ZnV$_2$O$_4$, which is similar to that shown by Maitra {\it et al.}\\

\underline{\textbf{MgV$_{2}$O$_{4}$ compound:}}\\

By using Table 5 and following the same procedure as in the case of ZnV$_{2}$O$_{4}$ we can get electron densities of V1 and V2 atoms for MgV$_{2}$O$_{4}$. The electron densities of V1 and V2 atoms are

\begin{eqnarray}
n_{1}=\frac{1}{(x^{2}+y^{2}+z^{2})^{2}}[0.23(x^{4}+y^{4})-0.19(xy)^{2}+0.08z^{4}+0.07(x^{2}+y^{2})z^{2}+0.70(yz)^{2}]
\end{eqnarray}
and
\begin{eqnarray}
n_{2}=\frac{1}{(x^{2}+y^{2}+z^{2})^{2}}[0.23(x^{4}+y^{4})-0.19(xy)^{2}+0.08z^{4}+0.07(x^{2}+y^{2})z^{2}+0.70(xz)^{2}]
\end{eqnarray}
respectively. It is clear from these equations that first four terms are same and hence provide same electronic distributions in the space. Last term is different which shows that the electronic distribution for V1 and V2 atoms are in ${yz}$ and ${xz}$ planes, respectively. suggesting the occupations of $d_{yz}$ and $d_{xz}$ orbitals in the global coordinate system. 

In MgV$_{2}$O$_{4}$ the $x$ and $y$ axes are making angles of 47.67 degree and 43.33 degree with two basal V-O bonds and $z$ axis is making an angle of 5.7 degree with the apical V-O bond. On rotating the $x$ and $y$ axes by 43.33 degree, the calculated electron densities for V1 and V2 atoms in local octahedral coordinate system become

\begin{eqnarray}
n_{1}&=&\frac{1}{(x^{2}+y^{2}+z^{2})^{2}}[0.07(x^{4}+y^{4})+0.79(xy)^{2}+0.08z^{4}-0.04xy(y^{2}-x^{2}) \nonumber \\ 
&&+0.4(xz-yz)^{2}+0.04(yz)^{2}+0.11xyz^{2}]
\end{eqnarray}
and
\begin{eqnarray}
n_{2}&=&\frac{1}{(x^{2}+y^{2}+z^{2})^{2}}[0.07(x^{4}+y^{4})+0.79(xy)^{2}+0.08z^{4}-0.04xy(y^{2}-x^{2}) \nonumber \\
&&+0.4(xz+yz)^{2}+0.04(xz)^{2}-0.11xyz^{2}] 
\end{eqnarray} 
respectively. The first four terms of Eqns. (11) and (12) are same and suggesting the same electron density in the space.  On dropping the first four terms from Eqns. (11) and (12) we get

\begin{eqnarray}
n_{1}=\frac{1}{(x^{2}+y^{2}+z^{2})^{2}}[0.4(xz-yz)^{2}+0.04(yz)^{2}+0.11xyz^{2}]
\end{eqnarray}
and
\begin{eqnarray}
n_{2}=\frac{1}{(x^{2}+y^{2}+z^{2})^{2}}[0.4(xz+yz)^{2}+0.04(xz)^{2}-0.11xyz^{2}] 
\end{eqnarray} 

On neglecting the last two terms which are almost four times smaller than the first term we get the same OO pattern for  MgV$_{2}$O$_{4}$ compound in the local octahedral coordinate system as observed in the ZnV$_{2}$O$_{4}$ compound. The small but different contribution from the last two terms are coming due to the fact that the VO$_6$ octahedron is not a regular one.\\ 

\underline{\textbf{CdV$_{2}$O$_{4}$ compound:}}\\

By using Table 8 and applying the same procedure as mentioned above, the calculated electron densities of V1 and V2 atoms, respectively for CdV$_{2}$O$_{4}$  are given by

\begin{eqnarray}
n_{1}=\frac{1}{(x^{2}+y^{2}+z^{2})^{2}}[0.23(x^{4}+y^{4})-0.23(xy)^{2}+0.07z^{4}-0.01(x^{2}+y^{2})z^{2}+0.82(xz)^{2}]
\end{eqnarray}
and
\begin{eqnarray}
n_{2}=\frac{1}{(x^{2}+y^{2}+z^{2})^{2}}[0.23(x^{4}+y^{4})-0.23(xy)^{2}+0.07z^{4}-0.01(x^{2}+y^{2})z^{2}+0.82(yz)^{2}]
\end{eqnarray}

Here also the first four terms in Eqns. (15) and (16) are same and last term is different. These equations also show the occupation of $d_{xz}$ and $d_{yz}$ orbitals in the global coordinate system as seen for the ZnV$_{2}$O$_{4}$ and MgV$_{2}$O$_{4}$ compounds. 

In CdV$_{2}$O$_{4}$ the $x$ and $y$ axes are making angles of 40.89 degree and 49.11 degree with two basal V-O bonds and $z$ axis is making an angle of 6.1 degree with the apical V-O bond. When we rotate the $x$ and $y$ axes by 40.89 degree, then the calculated electron densities for V1 and V2 atoms in local octahedral coordinate system become

\begin{eqnarray}
n_{1}&=&\frac{1}{(x^{2}+y^{2}+z^{2})^{2}}[0.06(x^{4}+y^{4})+0.79(xy)^{2}+0.07z^{4}-0.09xy(y^{2}-x^{2}) \nonumber \\
&&+0.44(xz+yz)^{2}-0.10(yz)^{2}-0.07xyz^{2}]
\end{eqnarray}
and
\begin{eqnarray}
n_{2}&=&\frac{1}{(x^{2}+y^{2}+z^{2})^{2}}[0.06(x^{4}+y^{4})+0.79(xy)^{2}+0.07z^{4}-0.09xy(y^{2}-x^{2}) \nonumber \\
&&+0.44(xz-yz)^{2}-0.10(xz)^{2}+0.07xyz^{2}]
\end{eqnarray}
respectively. Here also the first four terms in Eqns. (17) and (18) are same. Hence by dropping these terms we get
\begin{eqnarray}
n_{1}=\frac{1}{(x^{2}+y^{2}+z^{2})^{2}}[0.44(xz+yz)^{2}-0.10(yz)^{2}-0.07xyz^{2}]
\end{eqnarray}

and

\begin{eqnarray}
n_{2}=\frac{1}{(x^{2}+y^{2}+z^{2})^{2}}[0.44(xz-yz)^{2}-0.10(xz)^{2}+0.07xyz^{2}]
\end{eqnarray}

These equations also suggest the similar OO pattern for CdV$_{2}$O$_{4}$ compound in the local octahedral coordinate system as seen in the ZnV$_{2}$O$_{4}$ and MgV$_{2}$O$_{4}$ compounds.

Thus our studies on all the three compounds show exactly the same OO for all the compounds in the global coordinate system where $d_{xz}$ and $d_{yz}$ orbitals are mainly occupied by the neighboring V atoms as shown in the Fig. 2. In the local octahedral coordinate system the OO patterns for all the compounds show small deviation from $d_{xz}$+$d_{yz}$ and $d_{xz}$-$d_{yz}$ OO due to the presence of distorted VO$_6$ octahedron in the tetragonal structure of these compounds. Such deviation is expected to be important in understanding the magnetic properties of the compounds as the OO of a compound also decides the spin-ordering. Finally, it is important to note that quantification of the OO by including SOC will be an interesting studies on these compounds as the SOC is found to play an important role in deciding the magnetic properties of the compounds. The study of the effect of SOC on the OO is under progress and will be reported elsewhere.       
    
\section{Conclusions} 

   In conclusion we have studied the orbital ordering (OO) of the spinel vanadates AV$_{2}$O$_{4}$ (A $\equiv$ Zn, Cd and Mg), which is considered to be responsible for the low temperature structural transition from cubic to tetragonal phase in these compounds, by using density matrices approach to the OO. The GGA+U based \textit{ab initio} density functional theory has been used to calculate the density matrices of the vanadium atoms. In order to avoid extra complications we have not included the spin-orbit coupling for the valance electrons. In global (local octahedral) coordinate system anti-ferro orbital ordering has been observed in these compounds, where $d_{xz}$ and $d_{yz}$ ($d_{xz}$+$d_{yz}$ and $d_{xz}$-$d_{yz}$) orbitals are mainly occupied at the neighboring V sites.  
 
\acknowledgments {S.L. is thankful to UGC, India, for financial support.}


\pagebreak

\begin{table}
\caption{The density matrices correspond to 3${d}$ up electrons of V$_{1}$ and V$_{2}$ (in brackets) atoms for ZnV$_{2}$O$_{4}$ when {\it U} = 3 eV and {\it J} = 0.5 eV.}
\label{tab.4}
\begin{center}
\begin{tabular}{| c| c | c| c| c| c| c}
\hline
\backslashbox{$m1\downarrow$}{$m2\rightarrow$}&-2&-1&0&1&2
\\ \hline
-2&0.38(0.38)&$\approx 0$(0.02)&-0.05(0.05)&$\approx 0$(-0.06)&0.23(0.23)\\
\hline
-1&$\approx 0$(0.02)&0.47(0.47)&$\approx 0$(-0.01)&0.21(-0.21)&$\approx 0$(0.06)\\
\hline
0&-0.05(0.05)&$\approx 0$(-0.01)& 0.16(0.16)&$\approx 0$(0.01)&-0.05(0.05)\\
\hline
1&$\approx 0$(-0.06)&0.21(-0.21)&$\approx 0$(0.01)&0.47(0.47)&$\approx 0$(-0.02)\\
\hline
2&0.23(0.23)&$\approx 0$(0.06)&-0.05(0.05)&$\approx 0$(-0.02)&0.38(0.38)\\
\hline
\end{tabular}
\end{center}
\end{table}

\begin{table}
\caption{The density matrices correspond to 3${d}$ up electrons of V$_{1}$ and V$_{2}$ (in brackets) atoms for ZnV$_{2}$O$_{4}$ when {\it U} = 4 eV and {\it J} = 0.5 eV.}
\label{tab.1}
\begin{center}
\begin{tabular}{| c| c | c| c| c| c| c}
\hline
\backslashbox{$m1\downarrow$}{$m2\rightarrow$}&-2&-1&0&1&2
\\ \hline
-2&0.43(0.43)&$\approx 0$($\approx 0$) & -0.05(0.05) & $\approx 0$($\approx 0$) & 0.28(0.28)\\
\hline
-1&$\approx 0$($\approx 0$) &0.41(0.41)&$\approx 0$($\approx 0$)&0.32(-0.32)&$\approx 0$($\approx 0$)\\
\hline
0&-0.05(0.05)&$\approx 0$($\approx 0$)& 0.17(0.17)&$\approx 0$($\approx 0$)&-0.05(0.05)\\
\hline
1&$\approx 0$($\approx 0$)&0.32(-0.32)&$\approx 0$($\approx 0$)&0.41(0.41)&$\approx 0$($\approx 0$)\\
\hline
2&0.28(0.28)&$\approx 0$($\approx 0$)&-0.05(0.05)&$\approx 0$($\approx 0$)&0.43(0.43)\\
\hline
\end{tabular}
\end{center}
\end{table}

\begin{table}
\caption{The density matrices correspond to 3${d}$ up electrons of V$_{1}$ and V$_{2}$ (in brackets) atoms for ZnV$_{2}$O$_{4}$ when {\it U} = 5 eV and {\it J} = 0.5 eV.}
\label{tab.5}
\begin{center}
\begin{tabular}{| c| c | c| c| c| c| c}
\hline
\backslashbox{$m1\downarrow$}{$m2\rightarrow$}&-2&-1&0&1&2
\\ \hline
-2&0.44(0.44)&$\approx 0$($\approx 0$)&-0.09(0.09)&$\approx 0$($\approx 0$)&0.32(0.32)\\ \hline
-1&$\approx 0$($\approx 0$)&0.42(0.42)&$\approx 0$($\approx 0$)&0.36(-0.36)&$\approx 0$($\approx 0$)\\ \hline
0&-0.09(0.09)&$\approx 0$($\approx 0$)&0.16(0.16)&$\approx 0$($\approx 0$)&-0.09(0.09)\\ \hline
1&$\approx 0$($\approx 0$)&0.36(-0.36)&$\approx 0$($\approx 0$)&0.42(0.42)&$\approx 0$($\approx 0$)\\ \hline
2&0.32(0.32)&$\approx 0$($\approx 0$)&-0.09(0.09)&$\approx 0$($\approx 0$)&0.44(0.44)\\ \hline
\end{tabular}
\end{center}
\end{table}

\begin{table}
\caption{The density matrices correspond to 3${d}$ up electrons of V$_{1}$ and V$_{2}$ (in brackets) atoms for MgV$_{2}$O$_{4}$ when {\it U} = 3 eV and {\it J} = 0.5 eV.}
\label{tab.4}
\begin{center}
\begin{tabular}{| c| c | c| c| c| c| c}
\hline
\backslashbox{$m1\downarrow$}{$m2\rightarrow$}&-2&-1&0&1&2
\\ \hline
-2&0.40(0.40)&0.05($\approx 0$)&$\approx 0$($\approx 0$)&-0.06($\approx 0$)&0.22(0.22)\\
\hline
-1&0.05($\approx 0$)&0.48(0.48)&0.02($\approx 0$)&-0.18(0.18)&0.06($\approx 0$)\\
\hline
0&$\approx 0$($\approx 0$)&0.02($\approx 0$)& 0.19(0.19)&-0.02($\approx 0$)&$\approx 0$($\approx 0$)\\
\hline
1&-0.06($\approx 0$)&-0.18(0.18)&-0.02($\approx 0$)&0.48(0.48)&-0.05($\approx 0$)\\
\hline
2&0.22(0.22)&0.06($\approx 0$)&$\approx 0$($\approx 0$)&-0.05($\approx 0$)&0.40(0.40)\\
\hline
\end{tabular}
\end{center}
\end{table}

\begin{table}
\caption{The density matrices correspond to 3${d}$ up electrons of V$_{1}$ and V$_{2}$ (in brackets) atoms for MgV$_{2}$O$_{4}$ when {\it U} = 4 eV and {\it J} = 0.5 eV.}
\label{tab.2}
\begin{center}
\begin{tabular}{| c| c | c| c| c| c| c}
\hline
\backslashbox{$m1\downarrow$}{$m2\rightarrow$}&-2&-1&0&1&2
\\ \hline
-2&0.43(0.43)&0.02($\approx 0$)&$\approx 0$($\approx 0$)&-0.01($\approx 0$)&0.27(0.27)\\ \hline
-1&0.02($\approx 0$)&0.42(0.42)&0.04($\approx 0$)&-0.29(0.29)&0.01($\approx 0$)\\ \hline
0&$\approx 0$($\approx 0$)&0.04($\approx 0$)&0.21(0.21)&-0.04($\approx 0$)&$\approx 0$($\approx 0$)\\ \hline
1&-0.01($\approx 0$)&-0.29(0.29)&-0.04($\approx 0$)&0.42(0.42)&-0.02($\approx 0$)\\ \hline
2&0.27(0.27)&0.01($\approx 0$)&$\approx 0$($\approx 0$)&-0.02($\approx 0$)&0.43(0.43)\\ \hline
\end{tabular}
\end{center}
\end{table}

\begin{table}
\caption{The density matrices correspond to 3${d}$ up electrons of V$_{1}$ and V$_{2}$ (in brackets) atoms for MgV$_{2}$O$_{4}$ when {\it U} = 5 eV and {\it J} = 0.5 eV.}
\label{tab.4}
\begin{center}
\begin{tabular}{| c| c | c| c| c| c| c}
\hline
\backslashbox{$m1\downarrow$}{$m2\rightarrow$}&-2&-1&0&1&2
\\ \hline
-2&0.46(0.46)&0.01($\approx 0$)&0.03(-0.03)&$\approx 0$($\approx 0$)&0.31(0.31)\\
\hline
-1&0.01($\approx 0$)&0.43(0.43)&0.03($\approx 0$)&-0.34(0.34)&$\approx 0$($\approx 0$)\\
\hline
0&0.03(-0.03)&0.03($\approx 0$)& 0.19(0.19)&-0.03($\approx 0$)&0.03(-0.03)\\
\hline
1&$\approx 0$($\approx 0$)&-0.34(0.34)&-0.03($\approx 0$)&0.43(0.43)&-0.01($\approx 0$)\\
\hline
2&0.31(0.31)&$\approx 0$($\approx 0$)&0.03(-0.03)&-0.01($\approx 0$)&0.46(0.46)\\
\hline
\end{tabular}
\end{center}
\end{table}

\begin{table}
\caption{The density matrices correspond to 3${d}$ up electrons of V$_{1}$ and V$_{2}$ (in brackets) atoms for CdV$_{2}$O$_{4}$ when {\it U} = 3 eV and {\it J} = 0.5 eV.}
\label{tab.4}
\begin{center}
\begin{tabular}{| c| c | c| c| c| c| c}
\hline
\backslashbox{$m1\downarrow$}{$m2\rightarrow$}&-2&-1&0&1&2
\\ \hline
-2&0.42(0.42)&$\approx 0$($\approx 0$)&-0.08(0.08)&$\approx 0$(0.02)&0.27(0.27)\\
\hline
-1&$\approx 0$($\approx 0$)&0.43(0.43)&$\approx 0$(0.02)&0.29(-0.29)&$\approx 0$(-0.02)\\
\hline
0&-0.08(0.08)&$\approx 0$(0.02)&0.18(0.18)&$\approx 0$(-0.02)&-0.08(0.08)\\
\hline
1&$\approx 0$(0.02)&0.29(-0.29)&$\approx 0$(-0.02)&0.43(0.43)&$\approx 0$($\approx 0$)\\
\hline
2&0.27(0.27)&$\approx 0$(-0.02)&-0.08(0.08)&$\approx 0$($\approx 0$)&0.42(0.42)\\
\hline
\end{tabular}
\end{center}
\end{table}

\begin{table}
\caption{The density matrices correspond to 3${d}$ up electrons of V$_{1}$ and V$_{2}$ (in brackets) atoms for CdV$_{2}$O$_{4}$ when {\it U} = 4 eV and {\it J} = 0.5 eV.}
\label{tab.3}
\begin{center}
\begin{tabular}{| c| c | c| c| c| c| c}
\hline
\backslashbox{$m1\downarrow$}{$m2\rightarrow$}&-2&-1&0&1&2
\\ \hline
-2&0.43(0.43)&$\approx 0$($\approx 0$)&-0.06(0.06)&$\approx 0$($\approx 0$)&0.29(0.29)\\ \hline
-1&$\approx 0$($\approx 0$)&0.39(0.39)&$\approx 0$(0.01)&0.34(-0.34)&$\approx 0$($\approx 0$)\\ \hline
0&-0.06(0.06)&$\approx 0$(0.01)&0.18(0.18)&$\approx 0$(-0.01)&-0.06(0.06)\\ \hline
1&$\approx 0$($\approx 0$)&0.34(-0.34)&$\approx 0$(-0.01)&0.39(0.39)&$\approx 0$($\approx 0$)\\ \hline
2&0.29(0.29)&$\approx 0$($\approx 0$)&-0.06(0.06)&$\approx 0$($\approx 0$)&0.43(0.43)\\ \hline
\end{tabular}
\end{center}
\end{table}

\begin{table}
\caption{The density matrices correspond to 3${d}$ up electrons of V$_{1}$ and V$_{2}$ (in brackets) atoms for CdV$_{2}$O$_{4}$ when {\it U} = 5 eV and {\ J} = 0.5 eV.}
\label{tab.4}
\begin{center}
\begin{tabular}{| c| c | c| c| c| c| c}
\hline
\backslashbox{$m1\downarrow$}{$m2\rightarrow$}&-2&-1&0&1&2
\\ \hline
-2&0.36(0.36)&$\approx 0$(-0.09)&0.04(-0.04)&$\approx 0$(0.23)&0.20(0.20)\\
\hline
-1&$\approx 0$(-0.09)&0.51(0.51)&$\approx 0$(0.02)&-0.25(0.25)&$\approx 0$(-0.23)\\
\hline
0&0.04(-0.04)&$\approx 0$(0.02)&0.14(0.14)&$\approx 0$(-0.02)&0.04(-0.04)\\
\hline
1&$\approx 0$(0.23)&-0.25(0.25)&$\approx 0$(-0.02)&0.51(0.51)&$\approx 0$(0.09)\\
\hline
2&0.20(0.20)&$\approx 0$(-0.23)&0.04(-0.04)&$\approx 0$(0.09)&0.36(0.36)\\
\hline
\end{tabular}
\end{center}
\end{table}

\begin{figure}
\caption{Atomic arrangement of four V atoms namely V1, V2, V3 and V4 in the tetrahedron. V1(V3) and V2(V4) are orbitally inequivalent sites.}
\includegraphics{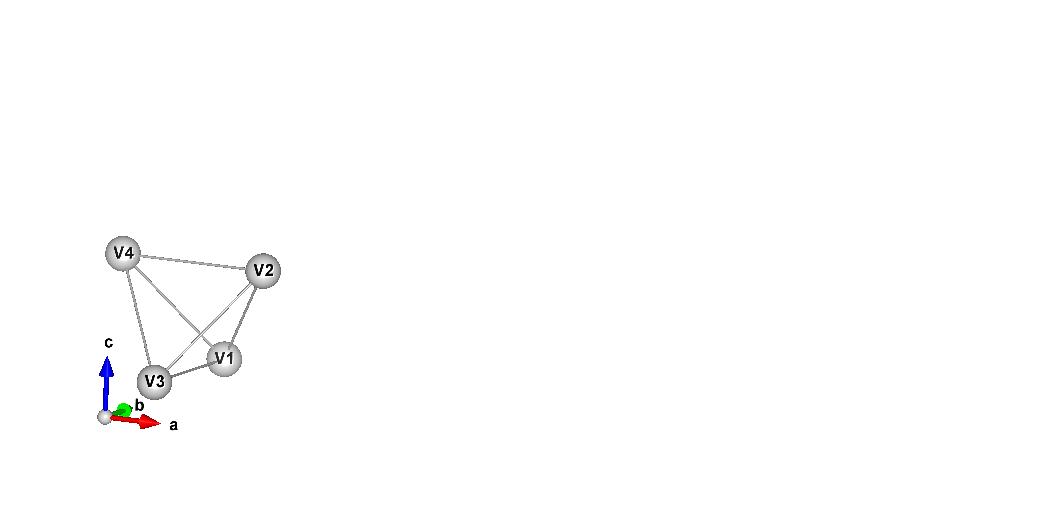}
\label{fig.1}
\end{figure}

\begin{figure}
\caption{Orbital ordering in AV$_{2}$O$_{4}$ where A = Zn, Mg and Cd at the octahedron level. {\it d$_{yz}$} and  {\it d$_{xz}$} orbitals are shown in blue and green, respectively.}
\includegraphics{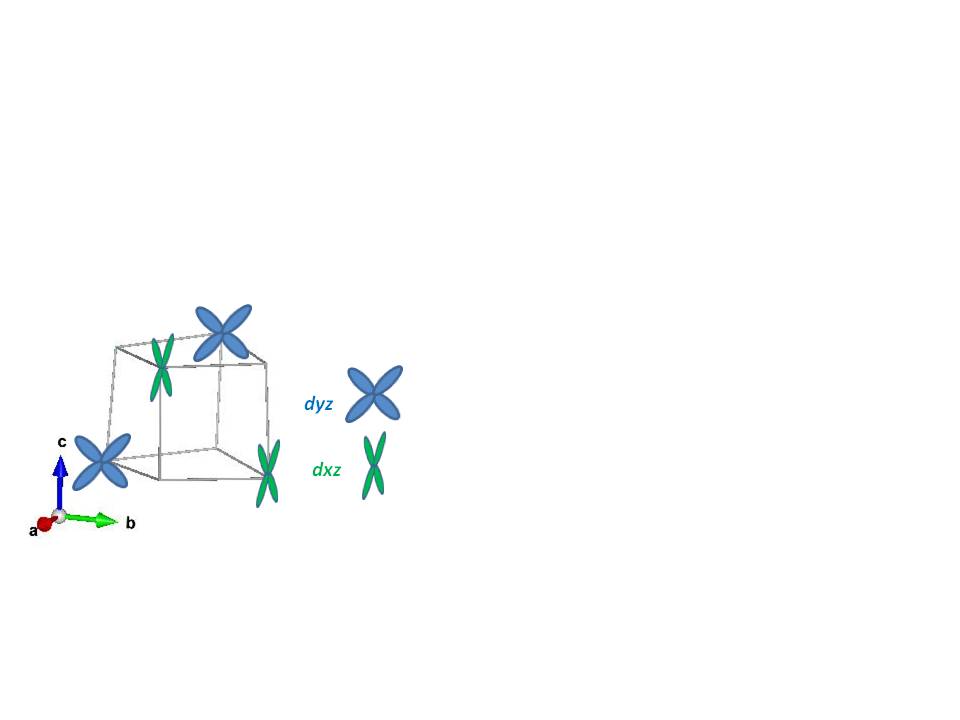}
\label{fig.2}
\end{figure}

\end{document}